\documentclass[12pt,preprint]{aastex631}

\usepackage{CJKutf8}
\usepackage{amssymb,amsmath}
\usepackage{graphicx,times}
\usepackage{natbib}
\usepackage{booktabs}
\usepackage{soul}

\newcommand    \Siflux       {$F_{\rm Si}$}
\newcommand    \Sflux        {$F_{\rm S}$}
\newcommand    \ratio        {$\gamma_{\rm S/Si}$}
\newcommand    \EFe          {$E_{\rm Fe}$}
\newcommand    \Bestshift    {$\Delta R_{\rm best}$}
\newcommand    \Vrsobs         {$v_{\rm{rs,obs}}$}
\newcommand    \Vrsej          {$v_{\rm{rs,ej}}$}

\shorttitle{Reverse shock in Cassiopeia A}
\shortauthors{Wu \& Yang}

\graphicspath{{./}{}}

\accepted{May 24, 2024}

\submitjournal{ApJ}

\begin{document}
\begin{CJK*}{UTF8}{gbsn}


\title{Reverse Shock Revisited in Cassiopeia A with \textit{Chandra}}

\author{Yin Wu (吴垠)}
\affiliation{Hunan Key Laboratory for Stellar and Interstellar Physics and School of Physics and Optoelectronics, Xiangtan University, Hunan 411105, China}

\author{X. J. Yang (杨雪娟)}
\affiliation{Hunan Key Laboratory for Stellar and Interstellar Physics and School of Physics and Optoelectronics, Xiangtan University, Hunan 411105, China}

\correspondingauthor{X. J. Yang}
\email{xjyang@xtu.edu.cn}

\begin{abstract}

Using data from the \textit{Chandra} X-Ray Observatory,
we revisited the reverse shock in the supernova remnant (SNR) Cassiopeia A.
Based on the spectroscopic of a series of annuli in the northwest (NW) and southeast (SE),
we get the radial profiles of the S/Si K$\alpha$ line flux ratio and Fe K$\alpha$ line centroid energy.
They both show monotonic increase, confirming that the Si- and Fe-rich ejecta are heated by the reverse shock.
The abrupt change of the S and Si line flux ratio is clearly observed in Cassiopeia A, 
leading to the determination of the reverse shock location 
($\sim$1.71$^{\prime}\pm$0.16$^{\prime}$ and $\sim$1.35$^{\prime}\pm$0.18$^{\prime}$ in the NW and SE, 
with respect to the central source).
By comparing the radial profiles of S and Si line flux, we find that the reverse shock is moving outward
in the frame of the observer, and the velocities are
$\sim$3950$\pm$210 km s$^{-1}$ and $\sim$2900$\pm$260 km s$^{-1}$ in the NW and SE, respectively.
In contrast, the velocities become $\sim$1150 km s$^{-1}$ (NW) and $\sim$1300 km s$^{-1}$ (SE) in the ejecta frame.
Our measured reverse shock velocities are quite consistent with those obtained from the
X-ray and/or optical images.
It therefore supplies a crosscheck of the accuracy for the two available methods to measure the
reverse shock velocity in SNRs.
Both the location and the velocity of the reverse shock show apparent asymmetry,
suggesting that the asymmetric explosion of the progenitor plays a key role in the interaction
between the reverse shock and the ejecta, ultimately shaping complex features observed in SNRs.

\end{abstract}


\section{Introduction}
\label{sect:intro}

During the supernova (SN) explosion, the fast-moving ejecta propagates into the ambient medium, generating a shock wave known as forward shock. The forward shock accelerates, compresses, and heats the gas, giving rise to X-ray emission \citep{Shklovskii1973}. In turn, the shocked gas decelerates the ejecta, giving rise to a shock wave moving back toward the expansion center, called reverse shock. It is usually a strong source of thermal X-ray emission \citep{Mckee1974} and an excellent astrophysical laboratory for understanding the SN explosion properties \citep[e.g.,][]{Sato2018}.

The X-ray line emission is closely related to the reverse shock heating history \citep{Mckee1974}. According to the reverse shock scenario, it passes through the outer layers of the ejecta earlier during the supernova remnant (SNR) evolution \citep{Chevalier1982,Truelove1999}. Therefore, the ejecta in the outer shells should be more highly ionized than those in the inner \citep{Chevalier2017}, leading to the expected radial variation of ionization age \citep[$\tau=n_et$, where $n_e$ is the number density of the electron and $t$ is the time since the shock heating,][]{Itoh1977,Vink2017}. However, it is difficult to obtain precise $\tau$ of the plasma in a specific region, since the observed plasma usually consists of many small-scale knots and clumps with various ionization states due to the projection effects \citep{Laming2003,Patnaude2007}. With this regard, it may be smoother to trace the reverse shock using the radial profiles of X-ray line emission properties, since such profiles average the effect of different ionization states over a larger area.

The radial profiles of Fe K$\alpha$ centroid energy ({\EFe}), metal line emissivity, and S/Si K$\alpha$ line flux ratio ({\ratio}) have been widely used to study the reverse shock in SNRs \citep[e.g.,][]{Gotthelf2001, Yamaguchi2014a, Lu2015}. The reverse shock may heat the ejecta in the outer region earlier, leading to a higher ionization degree and thus larger {\EFe} and {\ratio} \citep{Yamaguchi2014a,Lu2015}.
Furthermore, the ejecta heated by the reverse shock dominates the thermal emission in SNRs. The metal line emissivity may rise sharply near the current location of the reverse shock \citep{Gotthelf2001}. Therefore, these properties at different radii will provide an opportunity to study the reverse shock heating history and constrain the location of the reverse shock.

Cassiopeia A (Cas A) is an ideal object to study the reverse shock. With the age of $\sim$350 yr \citep{Thorstensen2001,Fesen2006b}, it is one of the most well-known young core-collapse (CC) SNR, arising from asymmetric Type IIb explosion \citep{Krause2008, Rest2011}. In this remnant, the reverse shock has not propagated into the most inner region, therefore it is possible to trace the time evolution of the reverse shock. Viewed in the X-ray, Cas A shows faint thin filaments with a radius of $\sim$2.5$^{\prime}$ \citep{Delaney2003} and a bright emission ring with a radius of $\sim$1.6$^{\prime}$ \citep{Gotthelf2001}. The filaments are identified as the current position of the forward shock front \citep{Gotthelf2001}, expanding at $\sim$5000 km s$^{-1}$ \citep{Delaney2003,Patnaude2009,Fesen2019,Vink2022}. The bright ring is composed of the ejecta heated by the reverse shock and is abundant in heavy elements, such as O, Ne, Mg, Si, Ar, Ca, and Fe \citep[e.g.,][]{Vink1996,Hughes2000,Delaney2004,Hwang2012}. According to the theoretical framework, nucleosynthesis in CC SNe occurs in a staged manner \citep{Woosley1995,Thielemann1996}. Fe-rich ejecta is expected to locate inside Si. However, spatially resolved spectroscopy shows that there is an overturn between the Fe-rich and Si-rich ejecta \citep{Hughes2000,Hwang2012,Tsuchioka2022}, which implies some strong central mechanism of explosion \citep[e.g.,][]{Sato2021}.

It is quite challenging to trace the reverse shock in young SNRs since the reverse shocked ejecta seems to be in knots or clumps \citep{Laming2003}. Nevertheless, there have been several works locating the reverse shock in Cas A \citep[e.g.,][]{Gotthelf2001,Helder2008,Arias2018}. In these works, the reverse-shocked region can be well separated from the outer forward shock. \cite{Gotthelf2001} estimate the reverse shock radius of 95$^{\prime\prime}\pm$10$^{\prime\prime}$ and mark the reverse shock as the sharp rise of emissivity. \cite{Helder2008} apply a deconvolution method for the nonthermal emission from the forward and reverse shock, and infers that the reverse shock is located at a radius of 115$^{\prime\prime}$ shifted to the west by 15$^{\prime\prime}$ with respect to the explosion center of Cas A \citep{Thorstensen2001}. More recently,  based on the internal free-free absorption in the radio, the location of the reverse shock was measured by \cite{Arias2018}. All these results are consistent, and the offset between the geometric and reverser shock center implies that the explosion asymmetry may affect the interaction between the reverse shock and the ejecta.

Moreover, the velocity of the reverse shock in Cas A remains to be studied. The theoretical model has predicted that the reverse shock is moving outward during the early phase of SNR \citep{Truelove1999}. According to the proper motion measurements \citep[e.g.,][]{Sato2018,Fesen2019,Vink2022} and three-dimensional (3D) hydrodynamic simulations \citep[e.g.,][]{Orlando2016,Orlando2022}, the reverse shock is moving outward in the northern and eastern of Cas A. The velocity in the frame of observer ranges between 2000 km s$^{-1}$ and 4000 km s$^{-1}$. However, the reverse shock in the southern and western of Cas A is stationary or even inward-moving with respect to the expansion center \citep[e.g.,][]{Delaney2004,Helder2008,Sato2018,Vink2022}, probably due to the explosion asymmetry \citep{Orlando2022}.

In this regard, we aim to trace the reverse shock in Cas A, and explore the effect of the asymmetric explosion on the shock structure.
Using archival \textit{Chandra} X-ray Observatory (\textit{Chandra}) data, we perform spectroscopy for a series of annuli in the northwest (NW) and southeast (SE) with abundant Si, S, and Fe line emissions \citep{Hwang2012}. The details of observation and data reduction are described in Section~\ref{sect:Obs}. In Section~\ref{sect:Res}, we show the radial profiles of {\EFe} and {\ratio}. Discussions in detail are presented in Section~\ref{sect:Dis}, regarding the heating history, proper motion and deceleration the reverse shock.
Finally, we give a summary in Section~\ref{sect:Sum}.

\section{Observation and data reduction} \label{sect:Obs}

\subsection{Region Selection} \label{sub:Reg}

This paper employs a series of observational datasets obtained by the \textit{Chandra} X-ray Observatory. 
The details of the data are listed in Table~\ref{Tab1}, and also contained in~\dataset[DOI: 10.25574/cdc.246]{https://doi.org/10.25574/cdc.246}. 
We selected these observations for the following reasons:
First, these observations have a relatively long exposure time ($>$30 ks) to ensure sufficient signal acquisition.
Therefore, we can do detailed spectral analysis for regions with limited areas.
Second, these observations span a time period ($\sim$20 year) as broad as possible to facilitate the measurement of the reverse shock velocity.
Finally, they have similar time intervals (4--6 year), making it possible to study the time evolution of the reverse shock.
We reprocessed these data using CIAO 4.13 \citep{Fruscione2006} and \textit{Chandra} Calibration Database (CalDB) 4.9.7.

\begin{table*}
\begin{center}
\caption{\textit{Chandra} observations}\label{Tab1}
\begin{tabular}{ccccc}
  \hline\hline\noalign{\smallskip}
ObsID. &  Date   & $\Delta t$ & Exposure & Dectector\\
       &         &  [yr]      &   [ks]   &          \\
  \hline\noalign{\smallskip}
114    & 2000-01-30  & 0     & 49.9  & ACIS-S  \\
4638   & 2004-04-14  & 4.23  & 164.5 & ACIS-S  \\
10936  & 2010-10-31  & 10.75 & 32.2  & ACIS-S  \\
14481  & 2014-05-12  & 14.28 & 49.4  & ACIS-S  \\
19606  & 2019-05-13  & 19.28 & 49.4  & ACIS-S  \\
  \noalign{\smallskip}\hline
\end{tabular}
\end{center}
\vspace*{-1.5em}
\end{table*}

\begin{figure*}
   \centering
   \includegraphics[width=16cm, angle=0]{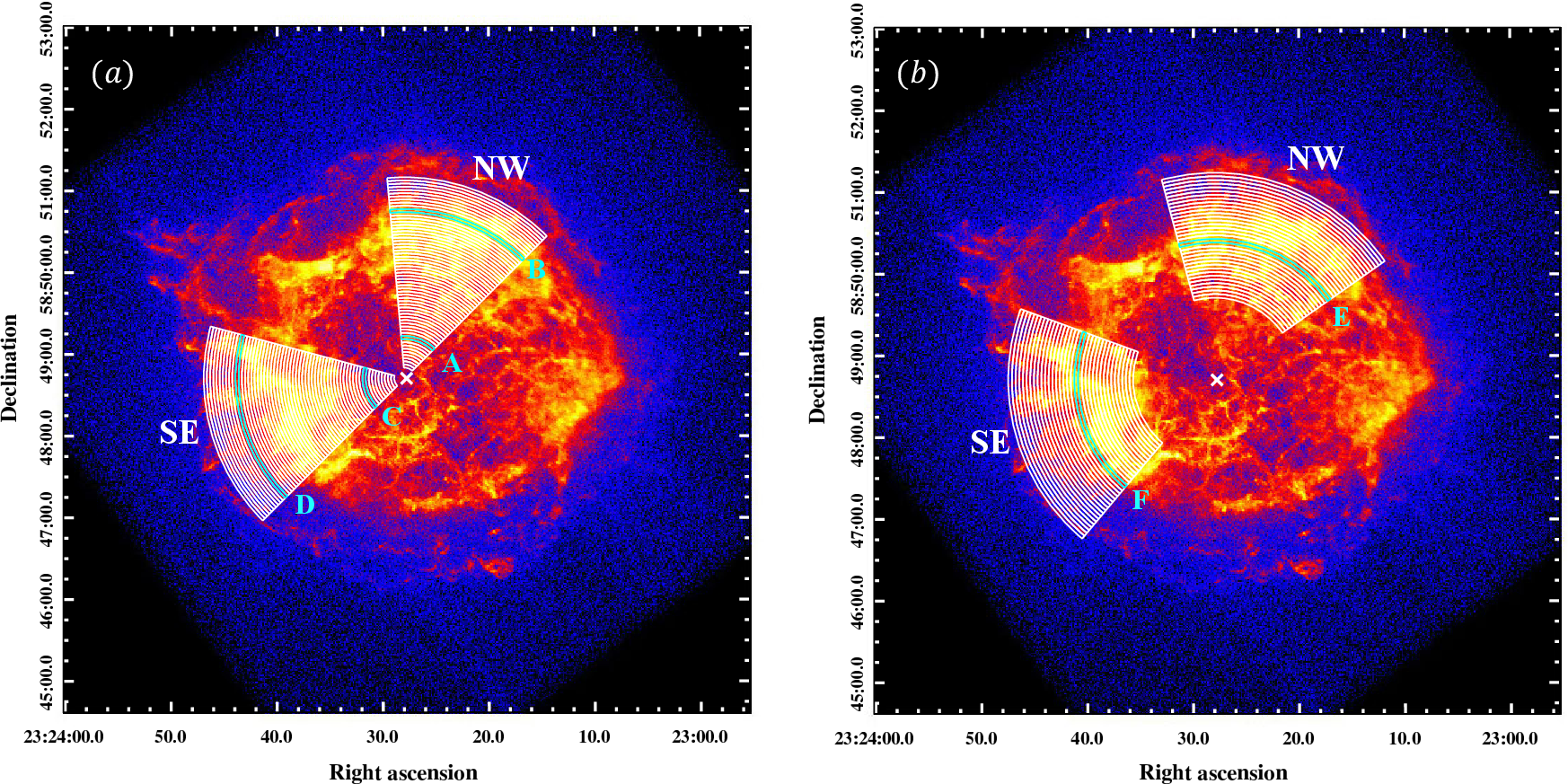}
   \vspace*{-0.5em}
   \caption{The 0.3--10.0 keV \textit{Chandra} X-ray image of Cassiopeia A in 2000 (ObsID. 114) with log brightness scale. (a): Selected 2$\times$60 annuli (white) in the NW and SE of Cas A for the study on {\ratio}. (b): Selected 2$\times$34 annuli (white) in the NW and SE of Cas A for the study on {\EFe}. The white cross marks the location of the central X-ray source in Cas A \citep{Fesen2006a}. The annuli marked as A--F (cyan) are characteristic regions for spectra demonstration in Figure~\ref{Fig2}.}
   \label{Fig1}
\end{figure*}

To obtain the radial profiles of {\ratio}, we extracted the 0.3--10.0 keV band spectra of 2$\times$60 annuli in the NW and SE, as illustrated in Figure~\ref{Fig1} (a). We focus on the Si-rich and S-rich regions with azimuth ranging from 315$^{\circ}$ to 5$^{\circ}$ in NW and from 75$^{\circ}$ to 135$^{\circ}$ in the SE \citep{Hwang2000,Hwang2012}, where 0$^{\circ}$ corresponds to due north. The width of the annulus is $\sim$ 0.039$^{\prime}$, while the minimum inner radius and the maximum outer radius are 0.1295$^{\prime}$ and 2.46$^{\prime}$, respectively.

Similarly, we extracted the spectra of the 4.0--8.0 keV spectra for 2$\times$34 annuli to study the radial profile of {\EFe}. Considering the different distribution of Fe with Si \citep{Delaney2004,Hwang2012}, we selected annuli with different parameters. The azimuth ranges from 305$^{\circ}$ to 15$^{\circ}$ in the NW and from 70$^{\circ}$ to 140$^{\circ}$ in the SE. The width of each annulus is 0.045$^{\prime}$. Fe line emission is deficient in the interior of Cas A, therefore we set the minimum inner radius to 1.005$^{\prime}$ and the maximum outer radius to 2.535$^{\prime}$. The selected regions are demonstrated in Figure~\ref{Fig1} (b).

To better demonstrate the effect of the explosion asymmetry on the reverse shock, we set the center for all our selected annuli at the
location of the central X-ray source, with RA $=$ 23$^h$ 23$^m$ 27.943$^s$ and DEC $=$ $+$58$^{\circ}$ 48$^{\prime}$ 42.51$^{\prime\prime}$ (J2000).
Considering that the explosion center is quite close to the point source, and its proper motion is relatively small \citep[$<$1 pixel=0.492$^{\prime\prime}$,][]{Sato2018},
we didn't adjust the position of our selected center.

\subsection{Spectra Fitting}
\label{sub:Spe}

\begin{figure*}
   \centering
   \includegraphics[width=12.25cm, angle=0]{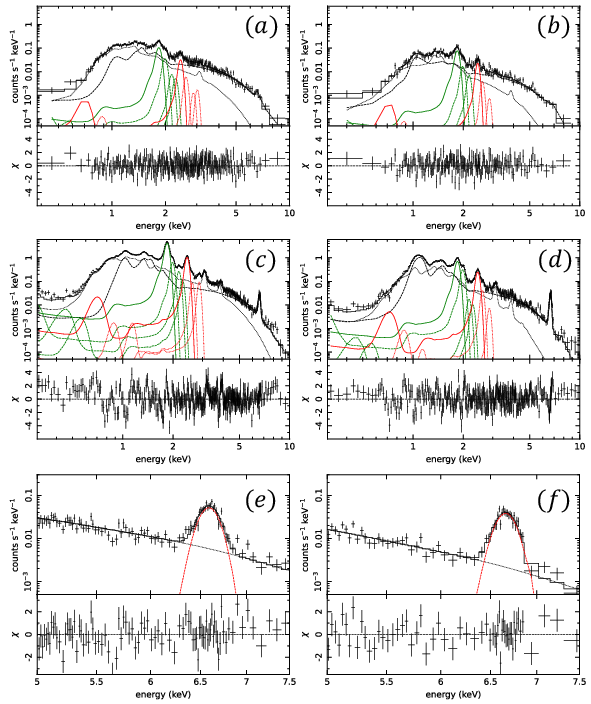}
   \vspace*{-0.5em}
   \caption{Examples of spectra from selected regions marked A--F in Figure~\ref{Fig1} for 2000 epoch (ObsID. 114). (a)--(d): The fitted spectra for the study on {\ratio}. The black dotted and dashed lines represent two non-equilibrium ionization components with the Si and S abundances fixed at 0. The green and red lines represent the Si and S line emission, respectively, and the prominent K$\alpha$ lines are marked as solid lines. (e) \& (f): The fitted spectra for the study on {\EFe}. The black and red lines represent the continuum component fitted by the power-law model and the Fe K$\alpha$ line fitted by the Gaussian model, respectively.}
   \label{Fig2}
\end{figure*}

We extracted spectra individually for each annulus at all the observation epochs, using \textit{specextract} in CIAO script. The background spectra were extracted from off-source regions. The spectra were analyzed with XSPEC version 12.12.0 \citep{Arnaud1996}.

To get {\ratio}, the 0.3--10.0 keV spectra of each annulus shown in Figure~\ref{Fig1} (a) at each epoch were fitted with the following
models just as those used by \cite{Lu2015}.
It includes a two-component non-equilibrium ionization (NEI) model \citep[VNEI in XSPEC,][]{Borkowski2001}
modified by photoelectric absorption (PHABS in XSPEC).
The free parameters are the hydrogen column density ($N_H$), temperature, emission measure (EM, $\int {n_e n_H} dV$),
ionization age $\tau$ and the abundance of O, Ne, Mg, Ar, Ca, Fe, and Ni for each VNEI component.
The Si and S abundance are set to zero, and eight Gaussian components are included to account for the Si and S emission lines,
where four for Si and four for S emission lines.
For Si, centroid energy of the K$\alpha$ line is frozen to be 1.86 keV, while the line width is set free.
The centroids for the Ly$\alpha$, He$\beta$, He$\gamma$ lines of Si are 2.006, 2.182, and 2.294 keV, respectively, and the
line widths are fixed to be 10$^{-5}$ keV.
Similarly, the S K$\alpha$ lines are set with free width of about 0.1 keV and centroid energy of about 2.45 keV.
For the the Ly$\alpha$, He$\beta$, and He$\gamma$ emission lines, the widths are fixed to 10$^{-5}$ keV
and centroid energies 2.623, 2.884, and 3.033 keV, respectively.
The {\ratio} are calculated with the K$\alpha$ emission lines, and the fluxes were obtained using \textit{cpflux} in XSPEC.

The 4.0--8.0 keV spectra of each annulus shown in Figure~\ref{Fig1} (b) at each observation epoch were fitted with a power-law model plus one Gaussian component to account for the continuum and Fe line emission, respectively. The {\EFe} for each annulus is thus given, with a 90\% confidence range.

Figure~\ref{Fig2} gives examples of the spectra as well as the fitting results. Figure~\ref{Fig2} (a)--(d) shows the spectra of the annuli marked A--D in Figure~\ref{Fig1} (a), as well as the fitting results. It is clearly shown that the spectra are different from region to region, not only the shape but also the strength of the emission lines. For regions A and C at the inner, the spectra are dominated by a featureless continuum with faint line emission and absence of the Fe K$\alpha$ line. Such regions possibly represent diffuse circumstellar material swept up by the projected forward shock, which is consistent with images in \cite{Delaney2004}. For regions B and D at the outer, the metal emission lines are quite strong. Meanwhile, they show different abundance patterns. Region B is Si- and S-rich, while region D is Fe-dominated.
Figure~\ref{Fig2} (e) \& (f) illustrate the spectra of regions E--F marked in Figure~\ref{Fig1} (b), respectively.
{\EFe} and the equivalent width are 6.595$^{+0.008}_{-0.008}$  keV \& 2.141 keV for region E
and 6.669$^{+0.010}_{-0.008}$ keV \& 4.391 keV for region F, respectively.
The equivalent widths indicate that region F exhibits a stronger Fe K$\alpha$ line emission compared to region E.
Considering region A \& C, region B \& D, region E \& F are at the same radius,
these different spectral features suggest asymmetric properties of the ejecta,
which might be associated with asymmetric SN explosion.

\subsection{Deprojection}
\label{sub:Deproject}

The surface flux of Cas A is the integration of the flux per unit volume. To eliminate the projection effect, we characterize the structure of Cas A as a spherical object
and obtain the radial profiles of the deprojected Si flux ({\Siflux}) and S flux ({\Sflux}).
Under the spherical assumption, the integral flux $F(r)$ is defined as \citep[e.g.,][]{Willingale1996,Helder2008,Lu2015}:
\begin{equation}{\label{eq1}}
F(r)=\int_r^R F_d(r')\dfrac{r'}{\sqrt{r'^2-r^2}}dr',
\end{equation}
where $R$ is the outer radius of the object, $F_d(r')$ the deprojected flux. The deprojection was performed from the outer radius to the inner radius by Equation~\ref{eq1} using the Lucy-Richardson technique \citep{Lucy1974}, and stopped when encountering the negative value.
We note here that although Cas A is very asymmetric, the deprojected radial profiles show little difference with the unprocessed ones.
The reason might be that the asymmetry of Cas A is mainly on a large scale, i.e., in different quadrants.
When it comes to the individual quadrant, e.g., SE, it is symmetric and therefore the projection is not significant.

\section{Results}
\label{sect:Res}

\subsection{Fe K Line Centroid Energy}
\label{sub:Fe}

\begin{figure*}
   \centering
   \includegraphics[width=15.5cm, angle=0]{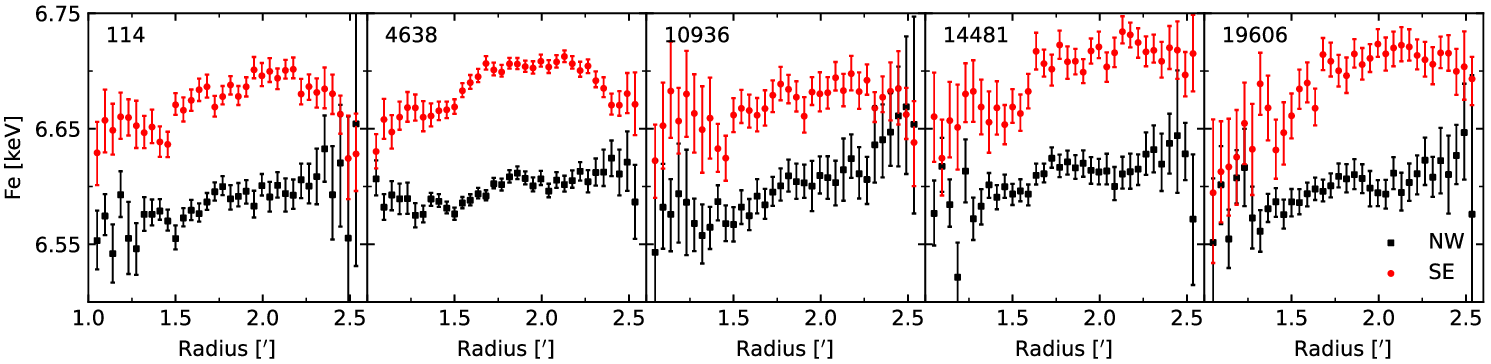}
   \caption{The radial profiles of {\EFe} at each epoch. The black square and red circle represent the fitted {\EFe} from the NW and SE, respectively. The error bars show the 90\% confidence range.}
   \label{Fig3}
\end{figure*}

Figure~\ref{Fig3} gives the radial profiles of {\EFe}. The best-fit {\EFe} is basically in the range of 6.55--6.7 keV, slightly larger than that from the whole remnant \citep[$\sim$6.6 keV,][]{Yamaguchi2014b}. Our fitting results show an overall larger {\EFe} in the SE than in the NW. The uncertainties of the {\EFe} can be quite large ($\sim$100 eV) in some selected regions ($R\gtrsim$2.3$^{\prime}$ in the NW and $R\lesssim$1.2$^{\prime}$ in the SE). We notice that in these regions the Fe K$\alpha$ lines are relatively weak. Therefore, we will focus on the region with 1.2$^{\prime}\lesssim R\lesssim$2.3$^{\prime}$ where Fe K$\alpha$ lines are relatively strong. Both in the NW and SE, {\EFe} increases monotonically with radius. We note that this monotonic increase can cover regions with a wide range of radii (1.5$^{\prime}\lesssim R\lesssim$2.3$^{\prime}$ in the NW and 1.3$^{\prime}\lesssim R\lesssim$2.1$^{\prime}$ in the SE).

\subsection{S/Si Line Flux Ratio and Location of the Reverse shock}
\label{sub:S/Si}

\begin{figure}
   \centering
   \includegraphics[width=15.5cm]{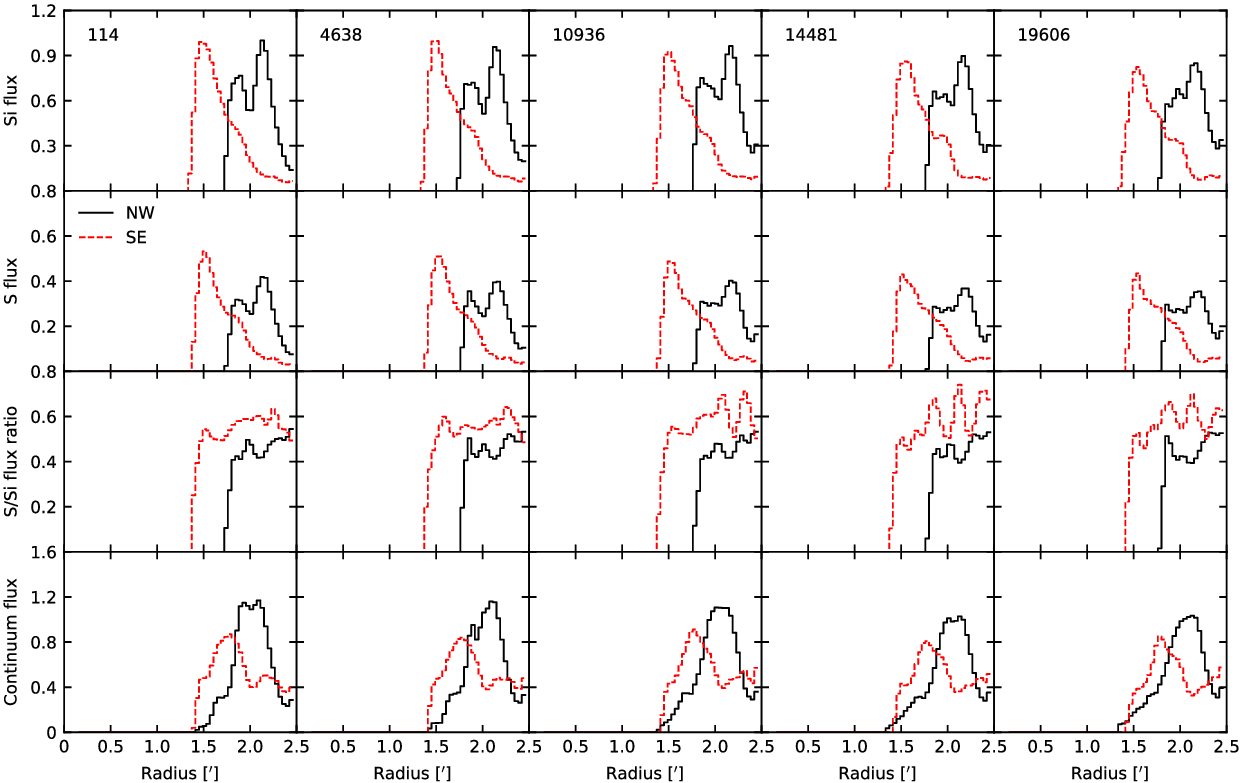}
   \caption{The radial profiles of the deprojected {\Siflux} (top panel), {\Sflux} (the second panel), {\ratio} (the third panel), and the thermal continuum flux (bottom panel) at each epoch from the NW (black solid line) and SE (red dashed line). All the fluxes are normalized by the maximum of the deprojected {\Siflux} in 2000 (ObsID. 114). The thermal continuum fluxes are represented by the fluxes in 0.85--1.5 keV plus 3.0--10.0 keV bands with the nonthermal component subtracted.}
   \label{Fig4}
\end{figure}

The radial profiles of the deprojected Si flux ({\Siflux}) and S flux ({\Sflux}) at each epoch are illustrated in the first two panels of Figure~\ref{Fig4}.\footnote{We note that after the deprojection, in the interior of the remnant ($R\lesssim$1.6$^{\prime}$ in the NW and $R\lesssim$1.3$^{\prime}$ in the SE), {\Siflux} and {\Sflux} become negative, therefore they are not shown in our profiles.}
The variations of {\Siflux} and {\Sflux} with radius are almost synchronous in the NW and SE.  As the radius increasing, {\Siflux} and {\Sflux} rise sharply and show apparent bump features. Both the location of the sharp rise and the shape of the bumps are different in the NW and SE.

The third panel of Figure~\ref{Fig4} gives the radial profiles of {\ratio} for the deprojected fluxes. {\ratio} increases almost monotonically with the radius, and we can also see a sharp rise ($R\sim$1.6$^{\prime}$ in the NW and $R\sim$1.3$^{\prime}$ in the SE). Remarkably, {\ratio} in the SE maintains a relatively high value ($\sim$0.6) in the outer radii of Cas A.

To estimate the current location of the reverse shock, the most concerning feature in the radial profiles of {\ratio} is the abrupt change. The region where the reverse shock has passed through may have a higher {\ratio} than the region where the reverse shock does not propagate \citep{Lu2015}. Thus, {\ratio} may change abruptly near the current location of the reverse shock.

\begin{figure}
   \centering
   \includegraphics[width=8cm]{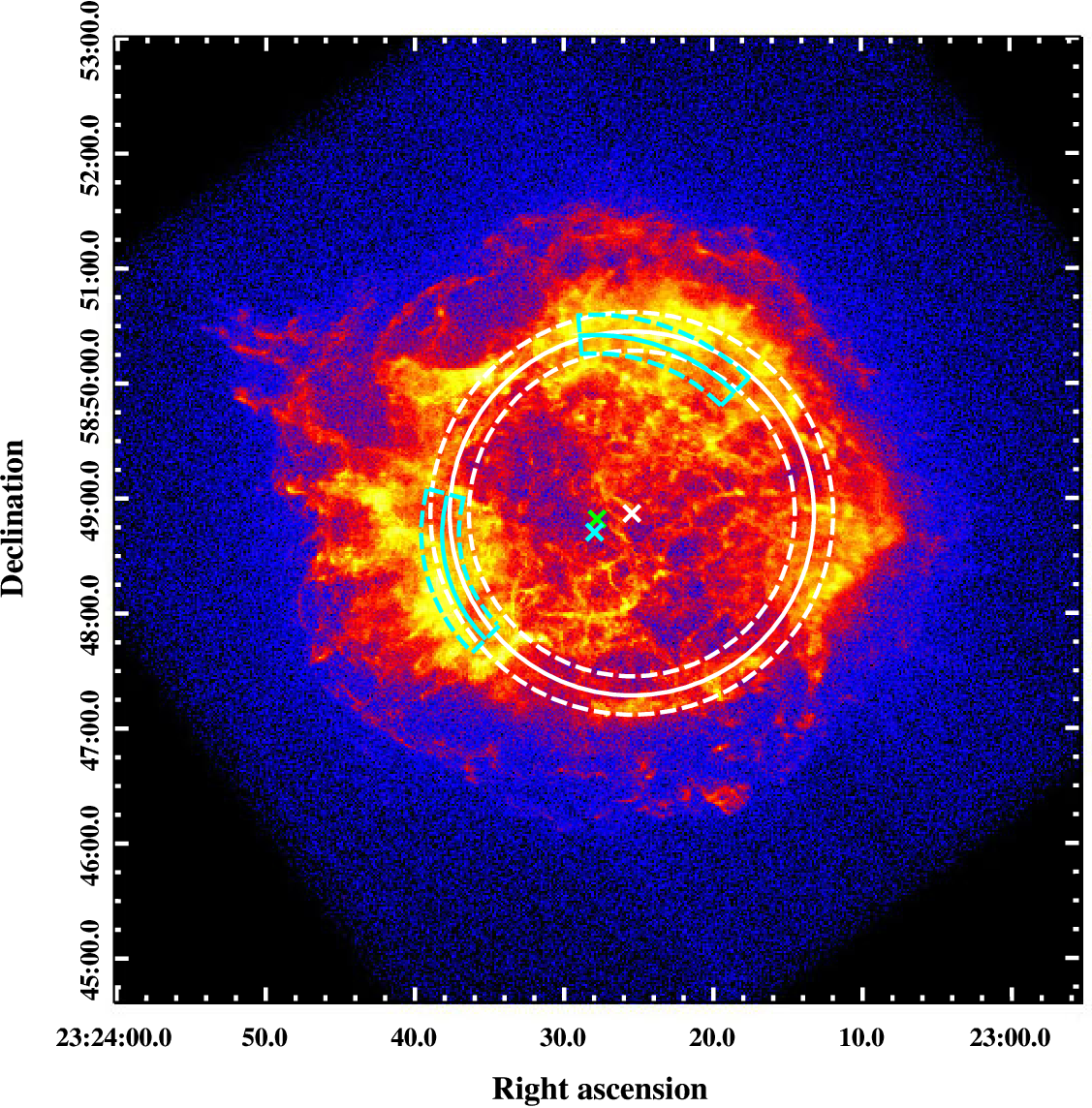}
   \caption{ The location of the reverse shock from our estimation (cyan) and \cite{Gotthelf2001} (white) for observation in 2000 (ObsID. 114). The solid annuli and circles mark the location of the reverse shock. The dashed annuli and circles represent the confidence range. Three crosses represent the possible explosion center \citep[green,][]{Thorstensen2001}, the central X-ray source \citep[cyan,][]{Fesen2006a}, and the geometric center of the reverse shock \citep[white,][]{Gotthelf2001}, respectively.}
   \label{Fig5}
\end{figure}

According to our radial profiles of {\ratio}, the reverse shock can be located. Since the location from the five observations are quite close, we estimated the average location at $R\sim$1.71$^{\prime}\pm$0.16$^{\prime}$ and $R\sim$1.35$^{\prime}\pm$0.18$^{\prime}$ in the NW and SE, respectively. The confidence range is given by the range of the sharp rise. Different reverse shock radii in the NW and SE suggest that the reverse shock may distribute asymmetrically in Cas A. \cite{Gotthelf2001} has estimated the location of the reverse shock from the sharp rise of the radio emissivity. The location of the reverse shock from our estimation and \cite{Gotthelf2001} are illustrated in Figure~\ref{Fig5}. We can see that our results are well consistent with theirs, although we use different approaches. Therefore, our method should be robust to estimate the location of the reverse shock.

\section{Discussion}
\label{sect:Dis}

\subsection{The Reverse Shock Heating History}
\label{sub:Heat}

We observe an overall radial monotonic increase of {\ratio} in the NW and SE, as shown in Figure~\ref{Fig4}.
According to \cite{Lu2015}, a higher {\ratio} can be related to a higher temperature and/or ionization age.
Since there is no abrupt change of temperature in the NW and SE \citep[e.g.,][]{Willingale2002,Yang2008,Hwang2012},
the most reasonable explanation of the radial monotonic increase is the radial increase of $\tau$.

Following \cite{Lu2015}, we investigate the contribution of $n_e$ and $t$ to the variation of $\tau$.
In the VNEI model, the normalization is in the units of $\dfrac{10^{-14}}{4\pi d^2}\int {n_e n_H} dV$ cm$^{-5}$ scaled with the EM,
where $d$ is the angular diameter distance to the source in the units of cm.
Assuming fully ionized H-dominant plasma ($n_e\approx$1.2$n_H$) \citep{Sato2021},
EM is proportional to $n_e^2$.
Therefore, the radial profile of $n_e$ can be simply estimated from that of the thermal continuum.
We calculate the thermal continuum fluxes in the 0.85--1.5 keV plus 3.0--10.0 keV bands
with the nonthermal contribution eliminated.
The nonthermal emission can be estimated by the extrapolation of the
4.2--6.0 keV spectra, since in this energy range it is nonthermal-dominated.
In the extrapolation, we use an index of 3.1 for all the spectra,
which is the average index of the entire remnant \citep{Helder2008,Sato2017}.
$N_H$ is set as 1.3$\times$10$^{22}$, a mean value for the absorption in the NW and SE \citep{Yang2008,Hwang2012}.
We also do a deprojection to the thermal continuum fluxes.

The results are plotted in the bottom panel of Figure~\ref{Fig4}. We can see that the radial profiles of the thermal continuum flux are distinctly different from {\ratio}. Thus, to explain the observed monotonic increase of {\ratio}, $t$ should play the key role, i.e., the ejecta in the exterior was heated earlier, implying that the reverse shock does pass through and heat the ejecta.

The radial profiles of {\EFe} also show monotonic increases. Similarly, we can deduce that the Fe-rich ejecta has been heated by the reverse shock. The reason is that a larger {\EFe} corresponds to a higher temperature and/or $\tau$ \citep{Yamaguchi2014a},
and no significant change in temperature is seen in these regions \citep[e.g.,][]{Willingale2002,Yang2008,Hwang2012}.

Though the monotonic increase is observed in the radial profiles of both {\ratio} and {\EFe}, the radius range is quite different.
Since the range of the monotonic increase of {\EFe} is far outside the average position of the reverse shock,
we suggest that this range may not be the precise position where the reverse shock heats the Fe-rich ejecta.
This is in good agreement with some special models to explain the interaction between the reverse shock and the ejecta,
such as the ``piston'' model \citep{Delaney2010} and the Fe-rich plumes \citep{Orlando2016,Orlando2021}.
The Fe-rich ejecta may be passed through and heated by the reverse shock in the interior of Cas A together with the Si-rich ejecta.
After the propagation of the reverse shock, the denser Fe-rich ejecta is slightly decelerated
and continues to move outward the remnant at a higher speed (at least higher than that of the Si-rich ejecta).
Finally, the Fe-rich ejecta will break through the average position of the reverse shock and occupy the outer
region of the Si-rich ejecta, resulting in the observed spatial inversion
between the Fe-rich and Si-rich ejecta \citep{Hughes2000,Hwang2012,Tsuchioka2022}.
Under this situation, the Fe-rich ejecta may be located outside the average radius of the reverse shock,
leading to larger radius ranges. Therefore, the observed large range of monotonic increase in the radial profiles of {\EFe} is acceptable.

Since the variation of {\Siflux}, {\Sflux}, and {\ratio} begin to be distinctly different in the exterior of the reverse shock,
the reverse shock may interact with the ejecta differently in the NW and SE. The relatively high {\ratio},
as well as the overall higher {\EFe}, suggests a higher ionization degree in the SE than in the NW.
Different ionization degrees may be related to different reverse shock actions caused by different
ejecta densities \cite[e.g.,][]{Tsuchioka2022}. The reverse shock may pass through the ejecta with lower
density more quickly and fully heat the ejecta, resulting in a higher ionization degree.
Accordingly, the ejecta density in the NW should be higher than that in the SE, which consists well with our proper motion measurements.
Nevertheless, due to the complex real density distribution of Cas A, higher quality data, such as the James Webb Space Telescope,
are needed for further understanding the interaction of the ejecta with the reverse shock \citep{Milisavljevic2024,Vink2024}.

\subsection{The Proper Motion of the Reverse Shock}
\label{sub:RSMotion}

We quantitatively estimate the proper motion ($\mu$) and velocity of the reverse shock by a $\chi^2$-test method described in \cite{Tanaka2020}.
We take the observation in 2000 (ObsID. 114) as the calibration standard, and then artificially move
the radial profiles of $F_{\rm Si}$ and $F_{\rm S}$ for the four other observations at different epochs.
Then we search for the best match shifts ({\Bestshift}) with $\chi_{\rm{min}}^2$ based on the $\chi^2$ defined as:
\begin{equation}{\label{eq2}}
\chi^2=\sum_k\dfrac{(F(i,k)-F(j,k))^2}{(dF(i,k))^2+(dF(j,k))^2},
\end{equation}
where $i$ and $j$ are the index of the epochs, $k$ is the index of the annuli, $F(i,k)$ and $F(j,k)$ are the fluxes for $k$ annulus at $i$ and $j$ epochs, respectively, $dF(i,k)$ and $dF(j,k)$ are the systematic errors of $F(i,k)$ and $F(j,k)$, respectively. {\Bestshift} represents the most possible shift, and the shifts satisfying $\chi^2\lesssim \chi_{\rm{min}}^2+1$ are given as the 1$\sigma$ confidence range.

\begin{figure*}
   \centering
   \includegraphics[width=16.5cm, angle=0]{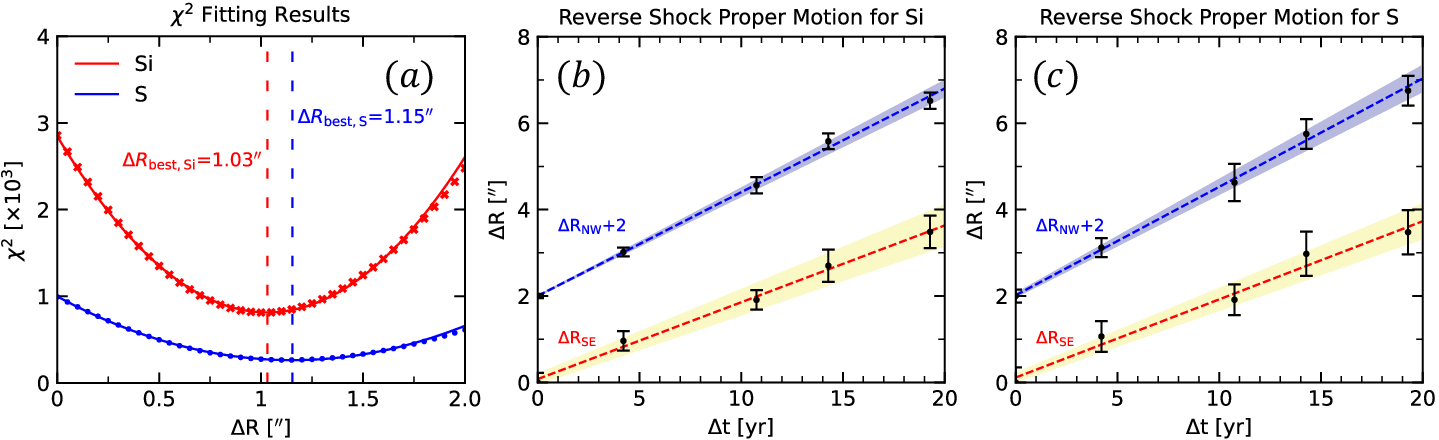}
   \caption{(a): $\chi^2$ as a function of shifts, derived from the radial profiles of {\Siflux} (blue) and {\Sflux} (red) for the NW in 2004 (ObsID. 4638) and 2000 (ObsID. 114). The scatters represent $\chi^2$ calculated for each shift. The curves are the results of a quadratic function fit. The dashed lines mark {\Bestshift}. (b): {\Bestshift} (scatters with error bars) versus time from observation in 2000 (ObsID. 114) as well the reverse shock proper motion fitting results (lines with the shaded area) for the radial profiles of {\Siflux} in the NW (blue) and SE (red) region. The error bars are given by the 1$\sigma$ confidence range of the best-fit shifts (with a factor of 5 for demonstration). The shaded areas are calculated by the standard deviations of the proper motion fitting. (c): Same as (b), but for the radial profiles of {\Sflux}. }
   \label{Fig6}
\end{figure*}

The {\Bestshift} are searched near the reverse shock, but avoid the plateau in the flux profiles.
According to the flux profiles and the estimated location of the reverse shock ($R\sim$1.71$^{\prime}$ in the NW and $R\sim$1.35$^{\prime}$ in the SE),
we set the range to $R\gtrsim$1.5$^{\prime}$ in the NW and $R\gtrsim$1.3$^{\prime}$ in the SE for our search.
We then using Equation~\ref{eq2} to calculate $\chi^2$ for each shift with a footstep of 0.05$^{\prime\prime}$.
And therefore the {\Bestshift} are obtained.
Figure~\ref{Fig6} (a) gives the example of the {\Bestshift} of observation in 2004 (ObsID 4638) with respect to that in 2000 (ObsID 114) for both Si and S in the NW.
{\Bestshift} are marked as dashed line, representing the most possible value of the reverse-shock motion between 2000 and 2004.
We can see that the {\Bestshift} for {\Siflux} and {\Sflux} are basically the same, indicating a common origin of Si and S in Cas A \citep{Woosley1995,Thielemann1996}.

We calculated the {\Bestshift} for all the epochs, and plotted them with respect to the time difference $\Delta t$.
They show a very good linear correlation, suggesting a common proper motion in all the epochs we used.
We therefore fit them with a linear function and give the slope, which represents the average proper motion.
Figure~\ref{Fig6} (b) \& (c) plot the fitting for {\Siflux} and {\Sflux}, respectively.
The fitting results are listed in Table~\ref{Tab2}.
The proper motion measured from the {\Siflux} and {\Sflux} are quite consistent
with each other. This is not surprising because Si and S are basically
co-located in Cas A (Yang et al. 2008).
For simplicity, we will adopt the average of the proper motion
measured from {\Siflux} and {\Sflux} as the final value, which
are also given in Table~\ref{Tab2}.
In the NW, it is about 0.245$\pm$0.013$^{\prime\prime}$ yr$^{-1}$, while
0.180$\pm$0.016$^{\prime\prime}$ yr$^{-1}$ in the SE.

With the proper motion measured, the velocity can thus be calculated from:
\begin{equation}{\label{eq3}}
v_{\rm{rs,obs}}=4740\times\left(\dfrac{\mu}{1^{\prime\prime}\ \rm{yr}^{-1}}\right)\times\left(\dfrac{d}{1\ \rm{kpc}}\right)\ \rm{km\ s^{-1}},
\end{equation}
where $v_{\rm{rs,obs}}$ is the velocity of the reverse shock in the frame of the observer,
$\mu$ the proper motion of the reverse shock, and $d$ the distance of Cas A adopted as 3.4 kpc \citep{Reed1995}.
$v_{\rm{rs,obs}}$ is $\sim$3950$\pm$210 km s$^{-1}$ in the NW,
and $\sim$2900$\pm$260 km s$^{-1}$ in the SE.
Based on the optical images between 2000 and 2004, \cite{Fesen2019} report a similar
{\Vrsobs} at 3500$-$3800 km s$^{-1}$ for the reverse shock front in the NW.
According to the comparison of X-ray images in \cite{Vink2022},
the reverse shock is moving outward at $\sim$4000 km s$^{-1}$ in the NW
(e.g., 4198$\pm$7 km s$^{-1}$ at azimuth$=$350$^{\circ}$) or
$\sim$3000 km s$^{-1}$ in the SE (e.g., 3056$\pm$32 km s$^{-1}$ at azimuth$=$90$^{\circ}$).
Within the confidence range, our results are well consistent with \cite{Fesen2019} and \cite{Vink2022}.
Our measurements thus provide a well crosscheck for the velocity of the
reverse shock obtained from the comparison of the optical and/or X-ray images.

To check whether the reverse shock is decelerated, we also calculate the proper motion within 2000 to 2010 and 2010 to 2019.
The results are listed in Table~\ref{Tab2}.
We can see that it basically decelerates by a factor of about 5\%.
Such deceleration have been reported by \cite{Delaney2003} theoretically.
They show that at the current evolution stage of Cas A, the deceleration factor is about 10\%.
Our observation is somehow consistent with the theoretical prediction, but slightly smaller.
Considering that we only have 20 years observations and the deceleration cannot be
very precisely constrained, future studies with larger time span would be helpful.

\subsection{The Deceleration to the Ejecta for the Reverse Shock }
\label{sub:RSDeceleration}

The ejecta will be decelerated when it passes through the reverse shock.
Such deceleration can be quantified more precisely if the effect of ejecta expansion is reduced,
and the reverse shock velocity in the ejecta frame {\Vrsej} will be a good indicator.
A larger {\Vrsej} means more deceleration \citep{Patnaude2009}.
We therefore make a rough estimate of {\Vrsej}.

According to the reverse shock dynamics \citep[e.g.,][]{Patnaude2009,Sato2018,Vink2022}, {\Vrsej} is defined as:
\begin{equation}
v_{\rm{rs,ej}}=|{v_{\rm{ej}}}-v_{\rm{rs,obs}}|,
\label{eq4}
\end{equation}
where $v_{\rm{ej}}$ is the velocity of the unshocked ejecta.
$v_{\rm{ej}}$ can be estimated from ${v_{\rm{ej}}}=R_{\rm{rs}}/t_{\rm SNR}$,
assuming the ejecta will be free expanding until encountering the reverse shock.
Here $R_{\rm{rs}}$ is the radius of the reverse shock, and $t_{\rm SNR}$ the age of the remnant.
Our estimated location of the reverse shock in Section~\ref{sub:S/Si} corresponds to
the angular distance of $\sim$1.7$^{\prime}$ in the NW and $\sim$1.4$^{\prime}$ in the SE,
taking the explosion center at RA $=$ 23$^h$ 23$^m$ 27.77$^s$, DEC $=$ $+$58$^{\circ}$ 48$^{\prime}$ 49.4$^{\prime\prime}$ (J2000) \citep{Thorstensen2001}.
The age of Cas A is adopted as $t_{\rm SNR}\approx$324 yr \citep{Fesen2006b}.
Then, $v_{\rm{ej}}$ is $\sim$5100 km s$^{-1}$ in the NW and $\sim$4200 km s$^{-1}$ in the SE.\footnote{
$v_{\rm{ej}}$ is estimated using the data of ObsID. 4638, because it has the longest exposure time
and can best constrain $v_{\rm{ej}}$.
}
Incorporated with the {\Vrsobs},
$v_{\rm rs,ej}$ are $\sim$1150 km s$^{-1}$ and $\sim$1300 km s$^{-1}$ in the NW and SE, respectively.
This suggests that the ejecta in the SE has suffered more deceleration from the reverse shock than in the NW.

We wondering if the different deceleration arises from different densities of the medium that the ejecta encounter.
The shock velocity and the density of the unshocked ejecta are related by the following equation given by \cite{Vink2022}:
\begin{equation}
\rho_{\rm{ej,0}}v_{\rm{rs,ej}}^2=\xi \rho_{\rm{csm,0}}v_{\rm{fs}}^2,
\label{eq6}
\end{equation}
where $v_{\rm{fs}}=m_{\rm{fs}}R_{\rm{fs}}/t$ is the forward shock velocity \citep[e.g.,][]{Patnaude2009},
$\rho_{\rm{ej,0}}$ the density of the unshocked ejecta,
$\rho_{\rm{csm,0}}$ the density of the unshocked circumstellar medium,
and $\xi$ a dimensionless parameter \citep[e.g., $\xi \approx$0.5 in][]{Vink2022}.
Since there is no significant change of the $v_{\rm{fs}}$ in the NW and SE \citep{Vink2022},
$v_{\rm{fs}}$ should not be responsible for different reverse shock velocities.
Therefore, different reverse shock velocities should be attributed to different density distribution
of ejecta and/or circumstellar medium.
A higher {\Vrsej} in the SE means that the ejecta density should be lower
(smaller $\rho_{\rm{ej,0}}$) or the ejecta may encounter denser circumstellar medium (larger $\rho_{\rm{csm,0}}$).

It has been suggested that different reverse shock velocities between the NW and SE might be associated
with different densities of the ejecta \citep{Fesen2019,Tsuchioka2022}.
When the reverse shock generates and passes through, the ejecta with lower density may suffer more deceleration.
Recently, considering the real density distribution of circumstellar medium,
a series of 3D hydrodynamic simulations have reproduced asymmetric reverse shock velocities in the NW and SE of Cas A \citep{Orlando2022}.
Their results suggest that the asymmetric velocities of the reverse shock are caused by the interaction
of the remnant with an asymmetric dense circumstellar shell.
However, their simulations do not reproduce all the sub-features in the velocity
profiles of the reverse shock measured by \cite{Vink2022}.
Therefore, the real density distribution of Cas A is very complex and needs further research.

\begin{table*}
\centering
\caption{Proper motions and velocities of the reverse shock in Cas A.
$\mu_1$ and $v_{\rm rs,obs,1}$ are derived from the radial profiles of {\Siflux}.
$\mu_2$ and $v_{\rm rs,obs,2}$ are derived from the radial profiles of {\Sflux}.
$\langle \mu \rangle $ and $\langle v_{\rm rs,obs} \rangle$ are the averages.
The confidence ranges are given by the standard deviations.}\label{Tab2}
\resizebox{.8\textwidth}{!}{
\begin{tabular}{lccccccc}
  \hline\hline\noalign{\smallskip}
       &  \multicolumn{3}{c}{$\mu_{1}$} & \multicolumn{3}{c}{$\mu_{2}$} & $\langle \mu \rangle$\\
       & \multicolumn{3}{c}{{[}$^{\prime\prime}$ yr$^{-1}${]}} & \multicolumn{3}{c}{{[}$^{\prime\prime}$ yr$^{-1}${]}}
       & {[}$^{\prime\prime}$ yr$^{-1}${]} \\
       \cmidrule(r){2-4} \cmidrule(r){5-7} \cmidrule(r){8-8}
Region &   2000--2019 & 2000--2010 & 2010--2019
       &   2000--2019 & 2000--2010 & 2010--2019 & 2000--2019\\
      \hline\noalign{\smallskip}
NW     &  0.239$\pm$0.014 & 0.238$\pm$0.021  & 0.226$\pm$0.029
       &  0.250$\pm$0.012 & 0.252$\pm$0.011  & 0.241$\pm$0.034  &  0.245$\pm$0.013\\
SE     &  0.178$\pm$0.018 & 0.188$\pm$0.021  & 0.184$\pm$0.018
       &  0.181$\pm$0.015 & 0.193$\pm$0.034 & 0.183$\pm$0.033  & 0.180$\pm$0.016\\
  \noalign{\smallskip}\hline\hline
       &  \multicolumn{3}{c}{$v_{\rm rs,obs,1}$} & \multicolumn{3}{c}{$v_{\rm rs,obs,2}$} & $\langle v_{\rm rs,obs} \rangle$\\
       &  \multicolumn{3}{c}{{[}km s$^{-1}${]}}  & \multicolumn{3}{c}{{[}km s$^{-1}${]}} &  {[}km s$^{-1}${]}\\
       \cmidrule(r){2-4} \cmidrule(r){5-7} \cmidrule(r){8-8}
Region &   2000--2019 & 2000--2010 & 2010--2019
       &   2000--2019 & 2000--2010 & 2010--2019 & 2000--2019\\
  \hline\noalign{\smallskip}
NW     &   3850$\pm$230 & 3840$\pm$340 & 3640$\pm$470
       &   4030$\pm$190 & 4060$\pm$180 & 3880$\pm$550 & 3950$\pm$210\\
SE     &   2870$\pm$290 & 3030$\pm$340 & 2970$\pm$290
       &   2920$\pm$240 & 3110$\pm$550 & 2950$\pm$530 & 2900$\pm$260\\
\hline\hline\noalign{\smallskip}
\end{tabular}
}
\end{table*}

\section{Summary}
\label{sect:Sum}

In this paper, we revisited the reverse shock in Cas A.
We perform a spatially resolved spectroscopy of Cas A with \textit{Chandra} observations
at different epochs.
We give the deprojected radial profiles of {\Siflux}, {\Sflux}, {\EFe} and {\ratio}, respectively.
{\ratio} increases monotonically with radius, indicating that the Si-rich ejecta has experienced the propagation of the reverse shock.
Similarly, the monotonic increase of the {\EFe} suggests reverse shock heating in the Fe-rich ejecta.
In the SE, overall larger {\EFe} and {\ratio} indicate higher ionization degree.
This might be related to different reverse shock actions. In the SE, the reverse shock may pass through the ejecta more quickly and fully heat the ejecta.

Based on {\ratio}, the reverse shock location is estimated to be $R\sim$1.71$^{\prime}\pm$0.16$^{\prime}$ in the NW and
$R\sim$1.35$^{\prime}\pm$0.18$^{\prime}$ in the SE with respect to the central X-ray source.
Within the confidence range, our estimated location is well consistent with previous results.
The reverse shock velocities in different directions are also given, by comparing the radial profiles of {\Siflux}
and {\Sflux} at different epochs by the $\chi^2$-test method.
In the frame of the observer, the reverse shock is moving outward at $\sim$3950$\pm$210 km s$^{-1}$
and $\sim$2900$\pm$260 km s$^{-1}$ in the NW and SE, respectively.
We marginally detected the reverse shock deceleration with a factor of $\sim$5\% from 2010--2019.
In the ejecta frame, the reverse shock velocities are $\sim$1150 km s$^{-1}$ and $\sim$1300 km s$^{-1}$ in the NW and SE, respectively.
The velocity in the ejecta frame suggests that the ejecta was more decelerated by the reverse shock in the SE,
and we suggest that
this should arise from the larger density that the ejecta encounters or the smaller density of the ejecta itself.

Overall, both the proper motion and heating history of the reverse shock show asymmetric distribution.
This should be closely related with the asymmetric explosion of the progenitor. 
During the explosion process, there may be an asymmetric mechanism that leads to asymmetric density distribution of the ejecta.
Or the explosion occurred in an asymmetric circumstellar medium.
Although we cannot conclude what the real situation is,
our results provide new insights on the effect of the asymmetric explosion in the progenitor of Cas A.

\bigskip

We thank F. J. Lu for helpful suggestions. We also thank the anonymous referee for thoughtful and helpful comments.
We acknowledge the use of data obtained by \textit{Chandra}.
The \textit{Chandra} Observatory Center is operated by the Smithsonian Astrophysical Observatory
for and on behalf of NASA.
This project is supported by the National Natural Science Foundation of China (NSFC 12122302 and 12333005).


\software{CIAO \citep[version 4.13; ][]{Fruscione2006}, XSPEC \citep[v12.12.0; ][]{Arnaud1996}. }

\end{CJK*}
\end{document}